\documentclass{PoS}

\usepackage{amsmath}
\usepackage{xspace}

\def \jpsi    {J/\psi}
\def \psip    {\psi(3686)}
\def \epem    {e^+e^-}
\def \D0      {D^{0}}

\def \Br   {\mathcal{B}}

\newcommand{\rom}[1]{\rm \uppercase\expandafter{\romannumeral #1\relax}}
\newcommand{\ee}{e^+e^-}

\newcommand{\dzero}{D^0}

\newcommand{\kpipi}{K^{+} \pi^{-} \pi^{+}}

\newcommand{\kpi}{K^{+} \pi^{-} }

\newcommand{\kpipipi}{K^{+} \pi^{-} \pi^{+} \pi^{-}}

\newcommand{\hhpee}{h (h') \ee}

\newcommand{\dptopipiee}{D^{+} \to \pi^{+} \pi^{0} \ee}
\newcommand{\dptokpiee}{D^{+} \to K^{+} \pi^{0} \ee}
\newcommand{\dptopiksee}{D^{+} \to K_{S}^{0} \pi^{+} \ee}
\newcommand{\dptokksee}{D^{+} \to K_{S}^{0}  K^{+} \ee}
\newcommand{\dztopipiee}{D^{0} \to \pi^{+} \pi^{-} \ee}
\newcommand{\dztokkee}{D^{0} \to K^{-} K^{+} \ee}
\newcommand{\dztokpiee}{D^{0} \to K^{-} \pi^{+} \ee}
\newcommand{\dztoksee}{D^{0} \to K_{S}^{0} \ee}
\newcommand{\dztoetaee}{D^{0} \to \eta \ee}
\newcommand{\dztoomegaee}{D^{0} \to \omega \ee}
\newcommand{\dztopizeroee}{D^{0} \to \pi^{0} \ee}

\newcommand{\gevcc}{\ensuremath{{\,\text{Ge\hspace{-.08em}V\hspace{-0.16em}/\hspace{-0.08em}}c^\text{2}}}\xspace}

\newcommand{\br}[1]{\mathcal{B}}

\newcommand{\mbc}{M_\mathrm{BC}}
\newcommand{\delE}{\Delta\emph{E}}

\newcommand{\JpsiDee}{$\jpsi \to \dzero \epem$~}
\newcommand{\PsipDee}{$\psip \to \dzero \epem$~}
\newcommand{\PsipLpee}{$\psi(3686)\rightarrow \Lambda_c^+ \overline{p} e^+ e^-$~}

\newcommand{\Dhhee}{$D\to\hhpee$~}
\newcommand{\Dhee}{$D^{+} \to h^{+}\ee$~}
\newcommand{\Dhees}{$D^{+} \to h^{-}e^{+}e^{+}$~}

\title{Searching for flavor changing neutral currents at BESIII}

\ShortTitle{Searching for flavor changing neutral currents at BESIII}

\author{\speaker{Dayong Wang}\thanks{Partially supported by Joint
    Funds of the National Natural Science Foundation of China (Grants
    No. U1832207) and by Ministry of Science and Technology(Grants
    No. 2015CB856700). } \\
  for the BESIII Collaboration\\
  \\
  State Key Laboratory of Nuclear Physics and
  Technology, Peking University, Beijing, China\\
  E-mail: \email{dayong.wang@pku.edu.cn}}

      \abstract{

        The Flavor Changing Neutral Current decays(FCNC) is forbidden
        at tree level in the Standard Model (SM) and could only
        contribute through loops.  Any direct observation beyond SM
        expectations could be a good probe of physics beyond
        SM. BESIII is a currently running tau-charm factory
        with the largest samples of on threshold charm meson pairs,
        directly produced charmonia and some other unique datasets at
        BEPCII collider. It has great potential to probe these FCNC
        decays from multiple channels. Here we review some of latest
        search results of FCNC decays from BESIII.  We present the
        latest results of searches for the decay of \JpsiDee,
        \PsipDee, \PsipLpee, \Dhhee, \Dhee, etc. Other related searches,
        together with the prospects and challenges with other channels
        and the future data are also discussed. }

      \FullConference{The International Conference on High Energy Physics - ICHEP2018\\
        4-11 July, 2018\\
        Seoul, South Korea}

\begin{document}
\section{Introduction}

BEPCII is the only currently running $\tau$-charm factory working at
the c. m. energy range of 2.0-4.6GeV, located at IHEP, Beijing. This
energy range has a lot of unique features which benefit greatly the
rich physics programs.  The BESIII detector has a geometrical
acceptance of $93\%$ of $4\pi$ and consists a small-celled,
helium-based main draft chamber, an electromagnetic calorimeter, a
time-of-flight system for particle identification, a muon chamber
system(MUC) made of about 1272 $m^2$ of resistive plate chambers
incorpared in the return iron of the superconducting solenoid with 1T
magentic field. More details of the detector are described in
Ref~\cite{bes3detector}.

The BESIII experiment has accumulated 1.3B $\jpsi$'s, 0.5B $\psip$'s
and $2.9 fb^{-1}$ at $\psi(3773)$, all of which are the largest data
sets in the world.  The clean environments and high luminosity at
BESIII provide ideal samples for searches for new physics(NP) beyond
standard model.

The Flavor Changing Neutral Current decays(FCNC) is forbidden at tree
level in the Standard Model (SM) due to the Glashow-Iliopoulos-Maiani
 mechanism~\cite{GIM}, and could only contribute through loops.
Any direct observation beyond SM expectations could be a good probe of
physics beyond SM.  BESIII has recently analysed several such FCNC
decays, which are reviewed in this paper, including the searches for
processes of \JpsiDee, \PsipDee, \PsipLpee, \Dhhee, etc. Charge
conjugation is always implied throughout.

\section{Search for \JpsiDee and \PsipDee~\cite{psidee} }
This analysis is performed with ($1310.6 \pm 7.2$) $\times$ $10^{6}$
$\jpsi$ events~\cite{jpsinum} and ($448.1 \pm 2.9$) $\times$ $10^{6}$
$\psip$ events~\cite{psipnum}, collected by BESIII.  In the SM, the
decay branching fraction for this kind of rare process is expected to
be of order $10^{-10}$ to $10^{-13}$. Some NP models could have
several of magnitudes higher decay rates, which could in the reach of
BESIII.

The $ \dzero $ signals are reconstructed through its three prominent
exclusive hadronic decay modes, $\kpi$, $\kpipi$, and $\kpipipi$,
which have relatively large branching fractions, and suffer from
relatively low background.  After requring one $ \dzero $ and one pair
of $ \ee $ reconstructed, the distributions of the invariant masses
from the three $\dzero $ meson decay modes are simultaneously fit with
unbinned maximum likelihood method for the $\jpsi$ and $\psip$
samples.

No $\dzero$ signals are observed, and we compute the upper limits(UL) on
the branching fraction at the 90\% C.L.\ using a Bayesian
method~\cite{PDG} with a flat prior,the correlated and un-correlated
systematic uncertainties are incorporated. The results are
$\Br(\jpsi\to\dzero \epem) < 8.5\times 10^{-8}$ and
$\Br(\psip\to\dzero \epem)< 1.4\times 10^{-7}$, respectively.  The limit
on $\Br(\jpsi\to\dzero \epem)$ is more stringent by two orders in
magnitude compared to the previous results, and the
$\Br(\psip\to\dzero \epem)$ is set for the first time.

\section{Search for \PsipLpee~\cite{psiplpee}}
\label{sec:psiplpee}

\begin{figure}[!htbp]
 \centering
 \includegraphics[width=0.3\textwidth]{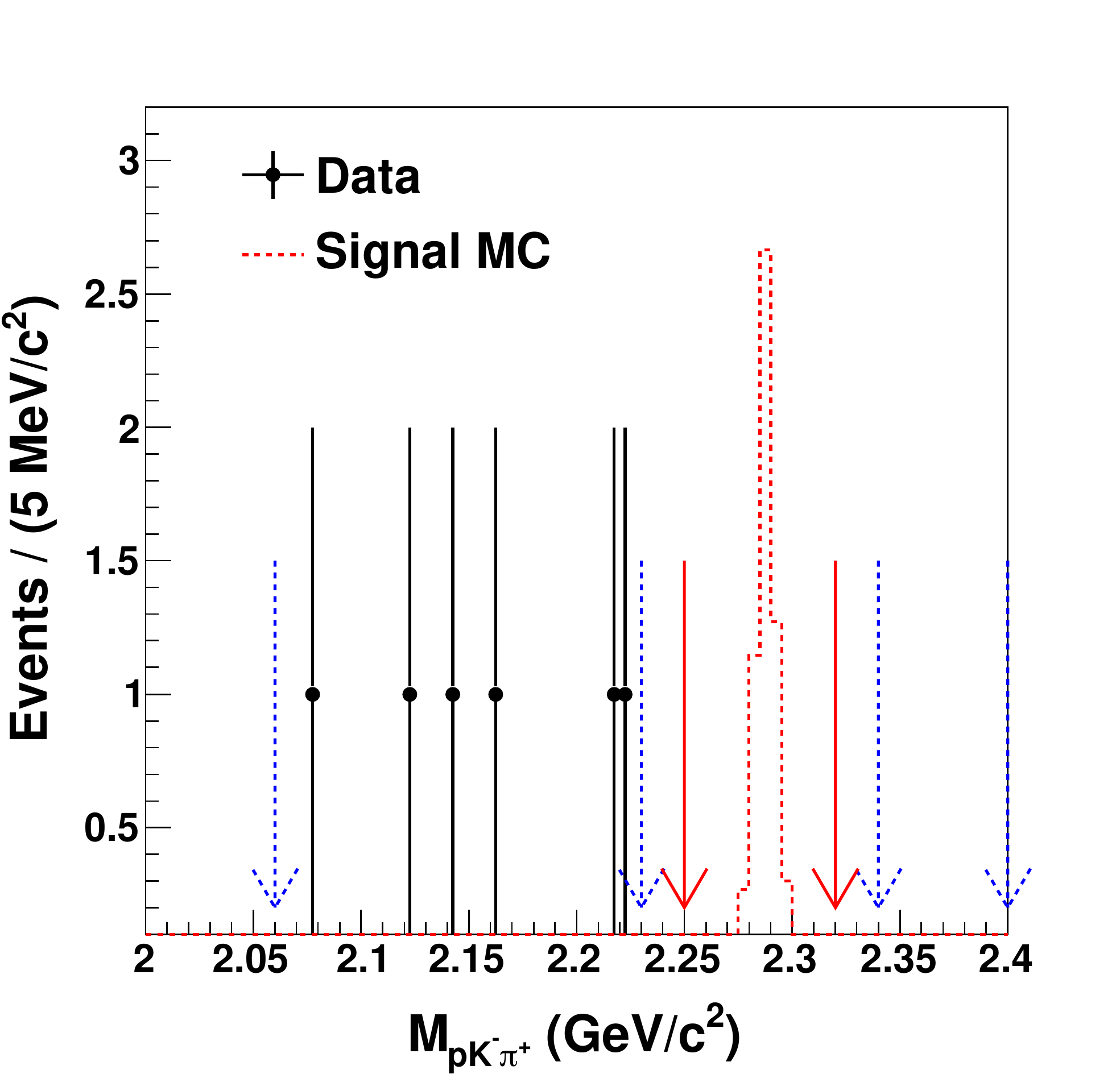}
 \includegraphics[width=0.45\textwidth]{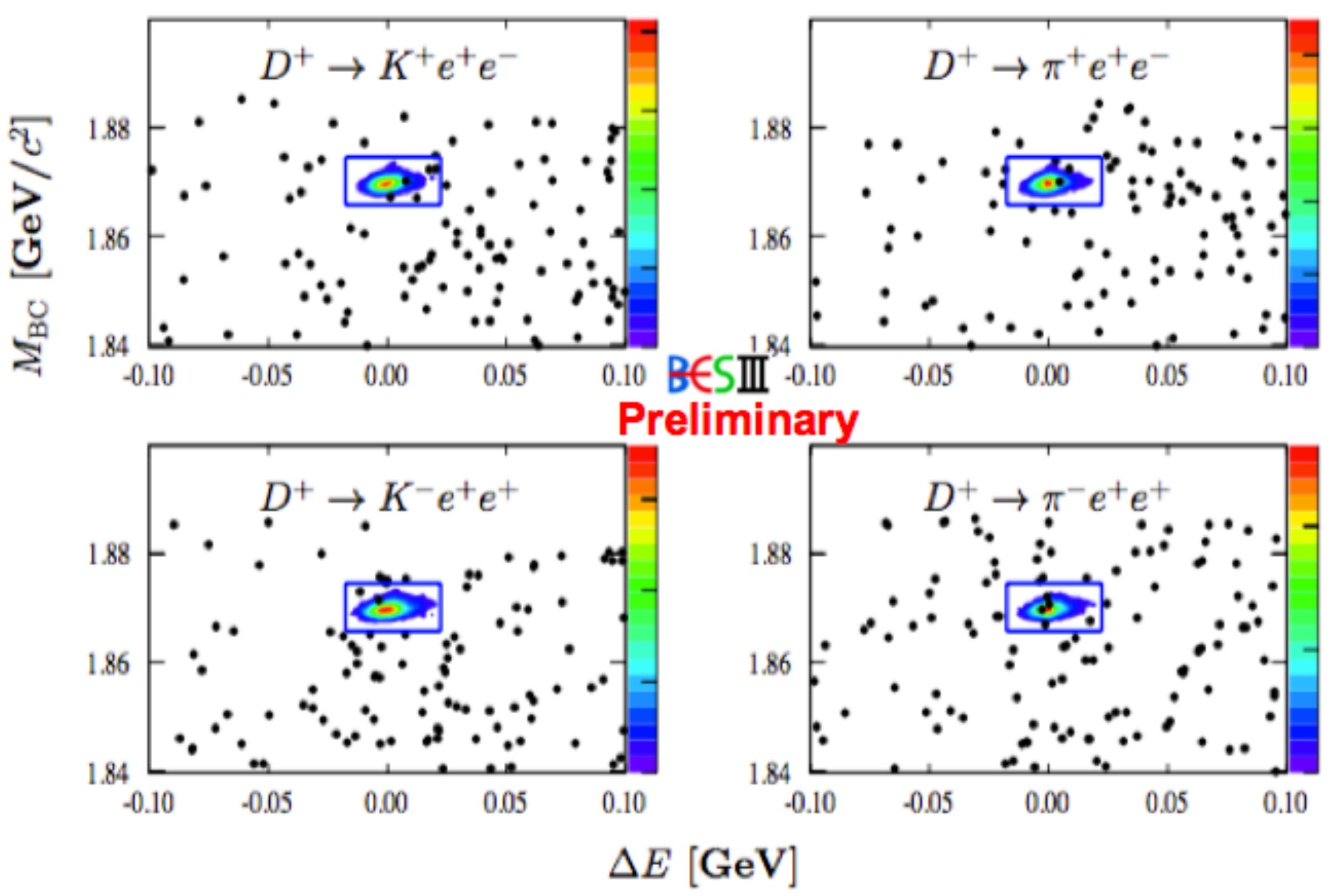}
\caption{Left: Distribution of $M_{pK^-\pi^+}$ for the data (dots with
  error bars) and signal MC sample (dashed histogram). Right: The
  scatter plots of  $\mbc$ versus $\delE$ of the accepted events in data, where
  the blue rectangle denotes the signal box, with the contour plots
  determined by MC simulation. }
   \label{fig:mpkpidata}
 \end{figure}

This analysis is performed with ($448.1 \pm 2.9$) $\times$ $10^{6}$
$\psip$ events~\cite{psipnum} collected by BESIII.  The
decay $\psi(3686)\rightarrow \Lambda_c^+ \overline{p} e^+ e^-$ with
$\Lambda_c^+ \rightarrow p K^- \pi^+$ is reconstructed with six
charged tracks with zero net charge.  The number of signal events is
determined by examining the $\Lambda^+_c$ signal in the
$M_{pK^-\pi^+}$ distribution, which is shown in
Fig.~\ref{fig:mpkpidata}(Left).

 No events survive within the signal region ranging from 2.25 to 2.32
 GeV/$c^2$.  The potential background in the signal region is
 estimated using events in the $M_{pK^-\pi^+}$ sideband regions to be
 1.5. We also estimate the number of background events to be zero
 using the inclusive MC sample and the data sample with
 $\sqrt{s}=3.773$~GeV. As no candidate events are found in the signal
 region, the estimated number of background events is determined to be
 $0\pm1.5$ events. The upper limit on the BF ($\mathcal{B}$) of the
 decay $\psi(3686)\rightarrow \Lambda_c^+ \overline{p} e^+ e^- + c.c.$
 is calculated to be $1.7\times 10^{-6}$. The result is within the
 expectations of the SM, and no evidence for new physics is found.

\section{Search for \Dhee and \Dhees}
\label{sec:dhee}

Using an $\ee$ collision sample corresponding to an integrated
luminosity of 2.93$\,\rm{fb}^{-1}$~\cite{luminosity1} collected with
the BES$\rom{3}$ detector at $\sqrt{s}$ = 3.773\,GeV, we search for
the FCNC decays \Dhee and the lepton-nubmer-violating decays \Dhees,
where $h$ are hadrons. The latter may be sensitive to NP models with
Majorana neutrinos. Fig.~\ref{fig:mpkpidata} (Right) shows the
resulting scatter plots of $\mbc$ versus $\delE$ of the accepted
events in data.

No signal excess is observed. As a result, we set the upper limits on
the branching fractions for these decays at a 90\% CL as:
$\Br(D^+ \to \pi^+ \ee) < 0.3\times 10^{-6}$,
$\Br(D^+ \to K^- e^+e^+ ) < 0.6\times 10^{-6}$,
$\Br(D^+ \to \pi^- e^+e^+ ) < 1.2\times 10^{-6}$,
$\Br(D^+ \to K^+ \ee) < 1.2\times 10^{-6}$. Compared limits from
previous experiments~\cite{PDG}, the first two results are
significantly improved than the previous restrictions and the other
two limits are comparable with the world best results. These are
BESIII preliminary results and formal publication will come later.

 \section{Search for \Dhhee~\cite{dhhee}}
\label{sec:dhhee}

Using the same dataset at$\sqrt{s}$ = 3.773\,GeV, we also perform a
search for the rare decays of $D\to \hhpee$, where $h^{(\prime)}$ are
hadrons. Double tagging(DT) method is used in the analysis.  For each
signal mode, $\delE_{\rm sig}$ is required to be within 3$\sigma$ of
the nominal value, and only the combination with the smallest
$|\delE_{\rm sig}|$ is kept. No significant excess over the expected
backgrounds is observed in $\mbc^{\rm sig}$ distributions of the
surviving events.

The ULs on the signal BFs at the 90$\%$ CL are determined.  The
maximal signal significance is $2.6 \sigma$, for $\dztokpiee$. Its BF
is expected to be dominated by the LD Bremsstrahlung and (virtual)
resonance decay contributions in the lower and upper regions, so we
divide the $M_{\ee}$ distribution into three regions and determine the
BFs in the individual regions.

\begin{table}[!hbtp]
\begin{center}
\scriptsize
\caption{Results of the ULs on the BFs for the investigated rare
  decays of \Dhhee at the $90\%$ CL, and the corresponding results in the PDG.
  \label{results}}
\begin{tabular}{lcc||lcc}
{Signal decays}        & $\br{}$($\times10^{-5}$)  &PDG~\cite{PDG}($\times10^{-5}$) &{Signal decays}        & $\br{}$($\times10^{-5}$)  &PDG~\cite{PDG}($\times10^{-5}$)  \\
\hline
$\dptopipiee$ & $<1.4$ & - & $\dztokkee$    & $<1.1$ & $ <31.5$ \\
$\dptokpiee$   & $<1.5$ & - & $\dztopipiee$  & $<0.7$  & $<37.3$ \\
$\dptopiksee$   & $<2.6$ & - & ${\dztokpiee}^{\dagger}$   & $<4.1$ & $<38.5$  \\
  $\dptokksee$  & $<1.1$ & - & $\dztopizeroee$  &$<0.4$ & $<4.5$ \\

$^{\dagger}$ in $M_{\ee}$ regions: & & & $\dztoetaee$  & $<0.3$ & $<11$  \\
$ \rm [0.00,~0.20)~\gevcc$  & $<3.0$ ($1.5^{+1.0}_{-0.9}$) & - &
                                                                 $\dztoomegaee$  & $<0.6$ & $<18$ \\
$\rm [0.20, ~0.65)~\gevcc$  & $<0.7$ & - & $\dztoksee$  & $<1.2$  & $<11$ \\
$\rm [0.65,~0.90]~\gevcc$  & $<1.9$ ($1.0^{+0.5}_{-0.4}$) & - \\
\hline
\end{tabular}
\end{center}
\end{table}

All these results are listed in Table~\ref{results}, and are all
within the SM predictions. For the four-body $D^+$ decays, the
searches are performed for the first time. The reported ULs of the
$D^0$ decays are improved in general by a factor of 10, compared to
previous measurements~\cite{PDG}. All the measured ULs on the BFs are
above the SM predictions, which include both LD and SD contributions.

\section{Summary and outlook}

BESIII has performed a series of searches for FCNC processes. Though
most of the upper limits are larger than the SM predictions, they may
help to discriminate the different new physics models or to constrain
the parameters in the different physics models.  Additionally, higher
statistics $\jpsi$, $\psip$ and D meason samples may help to improve
the sensitivity of the measurements. BESIII efforts with more related
channels and more coming data will be continued to further test the
standard model with higher precision in future.

\end{document}